\documentclass[12pt]{article}

\usepackage{a4wide}
\usepackage{cite}
\usepackage{amssymb}
\usepackage{amsmath}
\usepackage{epsfig}
\usepackage{eucal}

\newlength{\horizMargin}
\begin{document}


\titlepage

\begin{flushright}
DFF-435/12/06\\CERN-PH-TH/2006-246\\
December 2006
 \end{flushright}

\vspace*{1in}

\begin{center}
  {\Large \bf Infrared Sensitive Physics in QCD and in\\[3mm]
  Electroweak Theory}
\footnote{To appear in the volume {\it String Theory of Fundamental
    Interactions}, published in honour of Gabriele Veneziano in his 65-th
    birthday, M. Gasperini and J. Maharana editors ({\it Lecture notes in
    Physics}, Springer, Berlin/Heidelberg, 2007).}
  
  \vspace*{0.6in}
  
  Marcello Ciafaloni
  \footnote{Presently at CERN, Department PH-TH, Geneva}
  
  {\small
  \vspace*{0.5cm}
  {\it  Dipartimento di Fisica, Universit\`a di Firenze,
   50019 Sesto Fiorentino (FI), Italy} \\
  \vskip 2mm
  {\it  and INFN Sezione di Firenze,  50019 Sesto Fiorentino (FI), Italy}\\
  \vskip 6mm}
\end{center}


\bigskip
\begin{abstract}
\noindent
I recall the main ideas about the treatment of QCD infrared physics,
as developed in the late seventies, and I outline some novel applications
of those ideas to Electroweak Theory.
\end{abstract}

\newpage

\section{Infrared Sensitive Observables}
\label{sec:1}
The high energy physics of elementary particles, as described by the
Standard Model, gives particular emphasis to states constructed out of
{\it massless} partons or leptons, because of either the original
gauge symmetry, or of the QCD chiral symmetry. This
in principle introduces a number of problems because of the existence
of mass singularities in gauge theories -- that is, of infrared and
collinear divergences due to the initial or final states being
massless. Of course, physical states yield finite cross-sections
because of QCD confinement, or of Electroweak symmetry breaking, or of
QED coherent states. However, a remnant of the mass singularities of
the problem is that the cross-section, besides being dependent on
energy and momentum transfers of the process at hand, may also depend on
energy through large logarithmic variables, involving some infrared
sensitive mass parameters.

In QCD, avoiding large parameters is vital for the perturbative
description of hard processes, characterized by probe(s) with large
momentum transfer(s) $Q$ and by a supposedly small
coupling. Therefore, the cross-section must be infrared safe, i. e.,
sufficiently inclusive in order to cancel the mass singularities
according to the KLN and/or Bloch-Nordsieck (BN) theorems \cite{BN},
\cite{KLN}. As a consequence, fully inclusive processes are truly
perturbative, while the inclusive processes in which some partons of
virtuality $Q_0$ are looked at (in the initial or final state) show
anomalous dimensions \cite{DDT}. However, observables in which soft
emission is suppressed (e.g., at the boundary of the phase space) or
emphasized (e.g., of multiplicity type) are infrared sensitive
\cite{BCM}, and still contain parametrically large logarithms of
infrared origin, because of an incomplete cancellation of virtual
corrections with real emission.

The above observation raises a problem for quite interesting observables (like
$p_T$-form factors and jet multiplicity distributions), but indicates also how
to solve it because we know that the infrared behaviour is largely universal
due to the QED factorization theorem \cite{BN} and generalizations
thereof. This fact triggered, in the late seventies, a number of seminal
papers dealing with factorization of the collinear behaviour \cite{APV}, form
factor resummation \cite{ABCMV}, preconfinement \cite{AV}, jet evolution
\cite{KUV} and multiplicities \cite{BCM2}. It also appeared that one could
describe in full the final state \cite{BCM1} at the level of partons with
offshellness $Q_0$ much smaller than $Q$ but still large with respect to
$\Lambda$, the QCD scale, thus providing a ground for event generators
\cite{MW}.

All the above papers are largely based on factorization theorems for various
hard processes, and gradually introduce generalized renormalization group
techniques in order to predict the logarithmic dependence on the infrared
sensitive parameters at leading-logarithm anomalous dimension level, extended,
by further analysis \cite{CS}, to the subleading ones. The factorization
properties are in turn dependent on the cancellation of truly infrared
divergent contributions for all such processes, which requires a generalized
Bloch-Nordsieck theorem to be valid in QCD, as better established in the
eighties \cite{BBL}. In fact, the BN theorem states that a cross-section which
is inclusive over soft {\it final} states is also infrared safe, irrespective
of the fixed, possibly degenerate initial state. In this form, the theorem is
not automatically valid, because the nonabelian nature of QCD allows
degenerate initial states in a multiplet, which have different charges and
thus in general different cross-sections for the {\it same} momentum
configuration. This spoils the cancellation of virtual corrections with real
emission when summing over final soft states, unless an average over initial
colour is performed in order to restore the BN theorem. Fortunately, this
averaging is authomatic because of QCD confinement, which allows only colour
singlet asymptotic states.

The ideas above have been refined over the years in QCD, leading to an
approximate treatment of coherence effects by angular ordering in jet
evolution \cite{DKTM}, and to a more general treatment of subleading
logarithms in form factor calculations \cite{CT}. Recently, they have also led
to a new interesting development in Electroweak Theory. Na\"ively, one would
say that in the latter case the infrared structure is irrelevant because of
the spontaneously broken gauge symmetry, which provides a mass for weak bosons
and for fermions. However, with the advent of TeV scale accelerators, we shall
soon have access to energies which are much larger than the symmetry breaking
scale (say, the $W$ mass) which may act as {\it infrared} cutoff and thus give
rise to parametrically large infrared logarithms in the energy dependence, in
addition to the ones of collinear origin. That this is indeed the case was
first remarked in the late nineties \cite{CC} and soon applied to inclusive
observables \cite{CCC}. The failure of the BN theorem is due again to the
nonabelian nature of electroweak theory, where now {\it no averaging} over
flavour is possible, because the initial state consists of electrons, protons,
and so on, each of them having a nontrivial weak isospin charge. This also
means that double logarithms depending on the electroweak scale affect most
cross-sections which are apparently infrared safe, so that electroweak
radiative corrections are enhanced, sometimes comparable to QCD ones, and to
be carefully evaluated in a unified way.

My purpose in this note is to outline, in a few examples, how the novel ideas
of the seventies allow to understand the physics of large logarithms for both
QCD and Electroweak Theory, thus turning a potential problem into a powerful
tool. They also lead to a precise calculational framework for the logarithmic
energy dependence, for which I refer to the reviews already mentioned
\cite{BCM}, \cite{DKTM}, and to further dedicated papers \cite{CT}, \cite{C}.
 
\section{QCD Form Factors, Multiplicities, Preconfinement}
\label{sec:2}
\paragraph{Form Factors}
An early consequence of the understanding of infrared and collinear behaviours
in QCD was the remark \cite{ABCMV} -- \cite{BCM1} that observables where real
emission is suppressed are sensitive to the (square of) the partons' Sudakov
form factor. The latter is evaluated, at leading logarithmic level, by an
evolution equation in $ \mu^2 $ (the parton virtuality) which is derived by a
dispersive argument \cite{BCM},\cite{ABCMV}, or by applying \cite{EF} Gribov's
generalization of the Low theorem \cite{LG} as follows:

\begin {equation} 
\label{eq:sudakov}
 \frac{d\log F_a(Q^2,\mu^2)}{d\log \mu^2}~ =~ C_a \frac{\alpha_s(\mu^2)}{2\pi}
  \log(\frac{Q^2}{\mu^2})~,
\end {equation}
where $C_a= C_F,C_A$ is the Casimir charge of parton $a = q,g$. Note that
$\mu^2>Q_0^2$ plays the role of cutoff for an infrared divergent anomalous
dimension, so that $F_a$ shows an exponential suppression which, in the frozen
$\alpha _s$ limit, involves two logarithms per power of $\alpha _s$, one of
collinear type and the other of infrared origin. In the case of physical
observables, the cutoff on $\mu^2$ should be replaced by a parameter which
regulates real emission, like $Q^2/N$ for the parton PDFs at large moment
index $N$, or $1/B^2$ for impact parameter distributions. The outcome is the
characteristic large-$N$ dependence of PDFs for DIS and for the Drell-Yan
processes and the corresponding $p_T$-distributions.

For instance, the DIS structure function $F_N(Q^2)$ allows real
emission up to gluon momentum fraction $z<1/N$, and this regulates
the anomalous dimension of (\ref{eq:sudakov}) in the form

\begin{equation}
\label{eq:dis}
F_N(Q^2)~ \simeq~
\exp[-\frac{C_F}{\pi}\int_{Q_0^2}^{Q^2}\frac{d\mu^2}{\mu^2}\alpha_s(\mu^2)\log
Min(\frac{Q^2}{\mu^2},N)]~.
\end{equation}
We can see that the anomalous dimension becomes finite and of ~$logN$~
type for $\mu^2<Q^2/N$, while the ``exclusive'' limit is reached for
$N=Q^2/Q_0^2$, in which case Eq.(\ref{eq:dis}) reduces to
$F_q^2(Q^2,Q_0^2)$, where $Q_0$ is the minimal quark virtuality.

\paragraph{Multiplicities}
Actually, the idea underlying Refs. \cite{AV},\cite{BCM1} is to
describe outgoing hadronic jets in semi-inclusive form, at the level
of partons of virtuality $Q_0>\Lambda$, the decay products of the
latter being summed over. Here a problem of consistency arises,
because $Q_0$ is a somewhat arbitrary scale, and hadronic
distributions should be independent of it. Fortunately, two important
properties help. Firstly, multiplicity distributions show a {\it
factorized} $Q$-dependence with respect to the $Q_0$ dependence and, secondly,
{\it preconfinement} holds, namely the average mass of ``minimal''
colour singlets connected to a $q- \bar{q}$ pair is of order $Q_0$,
much smaller than $Q$. This means that jet evolution can be viewed in
two steps, a perturbative QCD evolution from $Q$ down to $Q_0$ (of
order $\Lambda$) and a hadronization process at scale $Q_0$. Thus, the
virtue of factorization and preconfinement is that the conversion into
hadrons does not affect the $Q$-dependence, and occurs at a much lower
scale.

Of course, the infrared analysis is essential in order to derive the
above properties. Factorization of multiplicity distributions is
argued for by resumming the double-log Feynman-$x$ dependence of jet
distribution functions in the soft region, which eventually leads to a
{\it finite} anomalous dimension with a singular $\alpha_s$-dependence
\cite{BCM2, BCM1} of type $\gamma_0\simeq \sqrt{\frac{N_c
\alpha_s}{2\pi}}$ \cite{M},\cite{EF}. Correspondingly, the average
hadronic jet multiplicity has the behaviour

\begin{equation}
\label{eq:multi}
\bar{n}(Q^2) \sim exp{\int_0^t dt \gamma_0(\alpha_s(t))} \simeq
exp\sqrt{\frac{2N_c}{\pi b}\log\frac{Q^2}{\Lambda^2}}~,
\end{equation}
and thus grows more rapidly than any power of $\log(Q^2/\Lambda^2)=t$.

The behaviour (\ref{eq:multi}) is remarkably different from the one of
QED radiation, essentially because of the gluon charge, implying that
the QCD jet evolution is a branching process, leading to a cascade,
rather than a bremsstrahlung process off one leg, as in
QED. Correspondingly, strong correlations of the final soft
partons are present, leading to an approximate KNO scaling of ``exclusive''
$n$-parton emission probabilities, which for a gluon jet have the form
\cite{BCM}
\begin{equation}
\frac{\sigma_n}{\sigma_{jet}}~\simeq~\frac{1}{\bar{n}}\exp[-\frac{1}{2}
(\log\frac{n}{\bar{n}})^2],~~~(n\ll \bar{n})~.
\end{equation}
This result shows that the the approximate proportionality of the $\sigma_n$s
in a gluon jet to the corresponding form factor (\ref{eq:sudakov}) still
holds, at double-log level, as for the electron in QED, but their relationship
to the average multiplicity (\ref{eq:multi}) - in the frozen $\alpha_s$ limit
- is quite different from QED because of the QCD cascade.

\paragraph{Preconfinement}
On the other hand, preconfinement \cite{AV} follows from a veto on the
possible final states which are allowed in the minimal colour singlets in
which, by definition, a $U(3)$ colour line connects a quark of offshellness
$Q_0$ to the corresponding antiquark. Because of factorization, and of the
veto, the inclusive mass distribution of minimal singlets being produced in a
jet of mass up to $Q$ is independent of $Q$ and is instead sensitive to the
quark form factor, as follows \cite{AV}, \cite{BCM1}

\begin{equation}
\label{eq:preco}
\frac{M^2 d\sigma}{\sigma_{jet} dM^2} \sim F_q^2(M^2, Q_0^2)~,
\end{equation}
so that its average mass is of order $Q_0$. Therefore, the conversion
of partons into hadrons can occur by an interaction of partons which
are close in phase space, leading to the so-called {\it local}
parton-hadron duality \cite{ADKT}, and to the possibility of building event
generators with relatively simple hadronization models
\cite{MW}, \cite{montecarlo}.

\section{Inclusive Electroweak Double Logarithms}
The infrared physics outlined above relies on the BN cancellation of
virtual and real emission singularities, which in QCD occurs because
of the colour averaging in the initial state, as remarked above.
Therefore, the form factor behaviour of type (\ref{eq:sudakov}) shows
up only if some veto uncovers the ``exclusive'' limit of the given
hard process. On the other hand, in Electroweak (EW) theory the BN
theorem {\it fails} because of the flavour charges of the accelerator
beams. For instance, the total cross-section for $e_+e_-$ annihilation
into hadrons is an infrared safe observable from the QCD standpoint,
but carries nevertheless EW double logarithms, embodied into an enhanced
effective coupling

\begin{equation}
\label{eq:coupling}
\alpha_{eff}(s) = \frac{\alpha_W}{4}(\log{\frac{s}{M_W^2}})^2~,
\end{equation}
which is of order $0.2$ in the TeV energy range and leads, therefore, to
sizeable corrections, of the same order as QCD ones. Besides the expected
collinear logarithm, the expression (\ref{eq:coupling}) carries an additional
one, of infrared origin, due to the violation of the BN theorem.

The analysis of such inclusive double logarithms \cite{CCC} involves form
factors of type (\ref{eq:sudakov}), where now $\mu^2$ is cutoff by the
EW scale $M_W^2 \simeq M_Z^2=M^2$ and the Casimir $C_a$ refers to the
isospin $I$ representation $a=I=0,1,...$ in the $t$-channel of the
lepton-antilepton overlap matrix. For instance, the combinations
$\sigma_{e_-\nu}\pm \sigma_{e_-e_+}$ correspond to $I=0$ $(I=1)$, so that

\begin{equation}
\label{eq:e+e-}
\sigma_{e_+e_-}(s,M^2) \simeq \frac{1}{2}(\sigma_0-\sigma_1F_1(s,M^2))
\simeq \frac{1}{2}(\sigma_0-\sigma_1\exp{(-2\frac{\alpha_{eff}(s)}{\pi})})~,
\end{equation}
where $\sigma_0$ corresponds to the isospin averaged cross section and has
therefore no double logarithms, while the antisymmetric combination $\sigma_1$
is damped by the $I=1$ form factor, with $C_1=2$. We note that, because of the
optical theorem, the inclusive form factor is not squared, though referring to
a physical cross-section in the crossed channel. Note also that in this
example $\sigma_1>0$, because the neutrino cross-section is larger, and
therefore the $\sigma_{e_+e_-}/\sigma_0$ ratio {\it increases} in the TeV
energy range towards its high-energy limit, which is provided by the flavour
average.

The above description can be generalized, by collinear factorization, to
single logarithmic level and to a generic overlap matrix involving leptons and
partons in the initial states, thus coupling the EW and QCD sectors of the
Standard Model. The result of this procedure is a set of evolution equations
in $\mu^2$ which are similar to the DGLAP equations \cite{DGLAP}, except that
evolution kernels exist in the channels with $I \neq 0$ also, and are infrared
singular or, in other words, depend on a logarithmic cutoff, much as in
Eq.(\ref{eq:sudakov}). For instance, in the evolution of lepton densities
$f_l$ and boson densities $f_b$, the $I=0$ evolution kernels coincide with the
customary DGLAP splitting functions $P_{ba}$, while the $I=1$ ones involve the
cutoff dependent virtual kernels

\begin{equation}
\label{eq:virtual}
P_f^V =~\delta(1-z)(- \log\frac{Q^2}{\mu^2} + \frac{3}{2}),~
P_b^V =~\delta(1-z)(-\log\frac{Q^2}{\mu^2}+\frac{11}{6}-\frac{n_f}{6})~.
\end{equation}
The corresponding evolution equations have the form
\begin{equation}
\label{eq:spin1}
- \frac{df^1_a}{d\log\mu^2}~=~\frac{\alpha_W}{2\pi} f^1_a P^V_a +~
  regular~~ terms~,
\end{equation}
and have been described in fully coupled form in \cite{CC1}. Here I just
notice that Eq.(\ref{eq:spin1}) shows a Sudakov behaviour similar to
(\ref{eq:sudakov}) and is consistent with Eq.(\ref{eq:e+e-}) after
taking into account the antilepton evolution, which doubles the
virtual kernel.

The presence of inclusive double logarithms in spontaneously broken gauge
theories remains an intriguing subject. It is mostly an initial state effect
and, as such, it is present for any final states of the same class (e.g.,
flavour blind) and strongly depends on the accelerator beams. Leptonic
accelerators maximize it, while hadronic ones (like LHC) provide some partial
average on the initial partonic flavours, thus decreasing it. But the effect
appears also if the flavour charges are looked at in the final state instead
of the initial state, for instance in gluon fusion processes in which some
$W$s are observed \cite{CC2}. Furthermore, the effect occurs whenever the
soft boson emission mixes several degenerate states having different hard
cross-sections. Non abelian theories have it because of the nontrivial
multiplets, but also a broken abelian theory shows it whenever the mass
eigenstates are not charge eigenstates \cite{CCC1}. An example of the latter
type is the mixing of the Higgs boson with the longitudinal gauge boson
occurring in a U(1) theory. The Standard Model shows both kinds of effects
and, given their magnitude in Eq.(\ref{eq:coupling}), I think that the coupled
evolution equations of parton-lepton distribution functions \cite{CC1} deserve
by now a quantitative study at the TeV scale.

Perhaps, the most important lesson to be learned from several decades of
investigation of infrared sensitive high-energy physics is that, even at the
level of hard processes, the fundamental interactions look much more
intertwined, due to the large time nature of asymptotic states which possibly
increases their effective couplings. By the same token, because of the large
times involved, factorization theorems are at work and allow a good
understanding of the infrared dynamics. It remains true, however, that a
unified treatment of all degrees of freedom is needed already at Standard
Model level -- that is, even before discovery of a possible short-distance
unification.

\section*{Acknowledgements}

It is a pleasure to thank Gabriele, as a collaborator and as a friend, for
sharing over many years and subjects the excitement of long discussions and,
sometimes, of real understanding. I also warmly thank old and new teams on
this subject, in particular Stefano, and Paolo and Denis, for various updates
of the picture presented here. I am finally grateful to the CERN Theory
Division for hospitality while this work was being completed, and to the
Italian Ministry of University and Research for a PRIN grant.

%
%

%
%



\end{document}